\documentclass[notitlepage,floats,aps,nofootinbib,twocolumn,prd,10pt,balancelastpage,longbibliography]{revtex4-1}
\usepackage{amsmath,amssymb,amsfonts}
\usepackage{mathtools}
\usepackage{hyperref,color}
\hypersetup{colorlinks,citecolor=blue,urlcolor=blue,linkcolor=blue}

\def\beq{\begin{equation}\begin{aligned}}
\def\eeq{\end{aligned}\end{equation}}
\def\OO{\mathcal{O}}
\def\dd{{\rm d}}
\newcommand{\mb}[1]{\boldsymbol{#1}}
\newcommand{\gs}{g_\star}
\newcommand{\gss}{g_{\star s}}

\newenvironment{tightequation}
{\setlength{\abovedisplayskip}{5pt}
 \setlength{\belowdisplayskip}{5pt}
 \begin{equation}}
{\end{equation}}

\begin{document}

\title{\vspace*{-7mm} Heavy Higgsino Dark Matter at the Scale of the Vanishing Higgs Quartic}

\author{James Unwin}
\affiliation{Department of Physics, University of Illinois Chicago,  IL 60607, USA}
\affiliation{Rudolf Peierls Centre for Theoretical Physics, University of Oxford, OX1 3NP, UK}

\begin{abstract}

The Standard Model Higgs quartic coupling runs to zero around $10^9-10^{12}$ GeV at $2\sigma$, suggestive of an ultraviolet boundary condition $\lambda\simeq0$ at this scale. In a well-motivated class of high-scale supersymmetric (SUSY) models, the quartic is automatically set to zero at tree level at the superpartner scale. The Higgsino can naturally emerge as the lightest SUSY particle, with its two neutral Weyl components sufficiently degenerate to form an effectively Dirac fermion with an unsuppressed vector coupling to the $Z$. For such Higgsino dark matter, direct detection requires a mass above $\sim10^{11}$ GeV, leading to a coincidence with the scale independently selected by the vanishing quartic. Such a heavy
Higgsino cannot freeze out with the observed relic density. However, if its mass exceeds the reheating temperature, the observed dark matter abundance can be reproduced through Boltzmann suppressed freeze-in. This scenario connects the dark matter mass to the measured Higgs mass, with a parameter space already constrained by LUX-ZEPLIN (LZ) and which will be decisively tested at DARWIN/XLZD. Finally, we consider PeV Higgsinos with Majorana gauginos of order $10^7$~GeV, permitting an inelastic Higgsino interpretation of the tentative LZ event while retaining the connection between the scalar superpartner scale and the vanishing Higgs quartic scale.

\end{abstract}

\maketitle

\section{Introduction}
\label{sec:intro}
\vspace{-3mm}

That the Standard Model (SM) gauge couplings converge under renormalisation group (RG) evolution has long been thought suggestive of a UV boundary condition $g_1=g_2=g_3$ set by new physics \cite{Georgi:1974sy}. Analogously, the observation that the Higgs quartic coupling runs to zero \cite{Degrassi:2012ry,Buttazzo:2013uya} is suggestive of a UV boundary condition
\begin{tightequation}
\lambda(\widetilde m)\simeq0.
\label{eq:boundary}
\end{tightequation}
From the measured Higgs mass, and assuming SM running, this identifies the scale: $\widetilde m\sim10^{9}-10^{12}~{\rm GeV}$.

The boundary condition eq.~(\ref{eq:boundary}) can arise at the scale of supersymmetry (SUSY) breaking. In the MSSM the quartic is not a free parameter but is inherited from the electroweak $D$-term. The quartic coupling vanishes if the light Higgs is aligned with the $D$-flat direction, which can be enforced exactly by symmetries in the Higgs sector  \cite{Hebecker:2012qp,Hebecker:2013lha,Ibanez:2012zg}, or if the $D$-term quartic itself is eliminated as occurs in theories with supersoft Dirac gauginos \cite{Unwin:2012fj}. In such models, the observed Higgs mass follows from a UV boundary condition on the Higgs potential at the intermediate superpartner scale, rather than from Weak scale naturalness or an anthropic explanation of the Higgs mass. The superpartners are therefore taken to lie near $\widetilde m\sim10^{9}-10^{12}$~GeV, as in high scale SUSY  \cite{Wells:2004di,Hall:2009nd}.

We highlight that these theories contain a natural dark matter candidate with distinctive and falsifiable phenomenology. The Higgsino mass is controlled by the $\mu$ parameter, introduced via the Giudice-Masiero or Kim-Nilles mechanisms \cite{Giudice:1988yz,Kim:1983dt}, and as such can be naturally lighter than the superpartner scale $\widetilde{m}$. For $|\mu|$ somewhat below the scalar and gaugino scale, the lightest supersymmetric particle (LSP) is an almost pure Higgsino, which is stable if R-parity (or matter parity) is conserved.

Importantly, the two neutral Higgsino degrees of freedom (DoF) are effectively degenerate. Majorana gaugino masses induce only a small neutral Higgsino splitting $\delta m_0 \sim m_Z^2/M_{1,2}$. For $M_{1,2}\sim10^{11}~{\rm GeV}$ then $\delta m_0$ is sub-keV, which is well below the $\OO(100~{\rm keV})$ inelastic threshold of direct detection experiments. Consequently, the Higgsino scatters as a Dirac electroweak state $\chi$ with an unsuppressed vector coupling to the $Z$, giving a neutron-level cross section $\sigma^n_Z\simeq7.4\times10^{-39}~{\rm cm}^2$, independent of mass for  $m_\chi\gg m_n$. Such a large cross section excludes Weak scale Higgsino dark matter, however, for heavier dark matter the event rate decreases with the local number density, and the Higgsino remains viable provided $m_\chi\gtrsim8\times10^{10}~{\rm GeV}$. Remarkably, this coincides with the upper end of the range independently selected by the vanishing Higgs quartic, and taken together we find the two requirements predict $m_\chi\simeq10^{11.4\pm0.5}$ GeV.

An electroweak state of this mass cannot be a conventional thermal relic. Partial-wave unitarity arguments imply that for a freeze-out relic the dark matter mass must lie below 
$m_\chi\lesssim 100$ TeV \cite{Griest:1989wd}. However, if the reheat temperature satisfies $T_{\rm rh}<m_\chi$, the observed abundance can be reproduced via Boltzmann suppressed freeze-in \cite{Giudice:2000ex,Cosme:2023xpa,Bernal:2025fcl}. The model presented here is a SUSY realisation of the ``minimal freeze-in dark matter'' \cite{Bernal:2026clv}. We show that the observed abundance is obtained for $T_{\rm rh}\sim m_\chi/27$, and the production rate never exceeds the Hubble rate, thus the Higgsinos never thermalise with the SM bath.
The resulting picture is economical and predictive, a symmetry of the UV Higgs sector explains the observed Higgs mass and the same UV scale provides a heavy, effectively Dirac Higgsino LSP. Direct detection requires its mass to lie near the same intermediate scale. Moreover, the relic abundance fixes the reheat temperature. This scenario will be discovered or excluded by next-generation direct detection experiments.

We first review the three SUSY mechanisms for realising eq.~(\ref{eq:boundary}). Then, section~\ref{sec:Higgsino} discusses the Higgsino LSP, the origin of $\mu$, and the neutral-Higgsino splitting. Section~\ref{sec:DD} examines direct detection implications. Section~\ref{sec:BSFI} studies the freeze-in abundance. Finally, in Section~\ref{sec:LZ}, motivated by the recent LZ high-recoil event \cite{LZ:2026axp}, we extend our discussion to consider a PeV inelastic Higgsino variant. We provide concluding remarks in Section~\ref{sec:conc}.

\section{Vanishing Higgs quartic as a  SUSY UV boundary condition}
\label{sec:UV}

In this section, we start by establishing the scale at which the quartic vanishes in the Standard Model, and review the SUSY mechanisms which give rise to the boundary condition
eq.~(\ref{eq:boundary}). We begin by fixing $\widetilde m$ from current measurements of $M_h$, $M_t$ and $\alpha_s$ (where we use upper case $M_i$ to indicate a pole mass). The SM scalar potential (or equivalently its supersymmetric extensions with all of the superpartners and heavy Higgses integrated out) is given by
\beq
V(H)=-\mu_{\rm SM}^2 |H|^2 + \lambda|H|^4.
\label{VSM}
\eeq
Thus $m_h^2=2\lambda v^2$ with $v\simeq246$~GeV and $\lambda(m_t)\simeq0.126$. Under SM running $\lambda$ decreases as one runs into the UV, crossing zero at an intermediate scale. To estimate this scale, we take the NNLO fit of~\cite{Buttazzo:2013uya} for identifying the instability scale $\Lambda_*$ of the SM Higgs potential
\beq
\log_{10}\frac{\Lambda_*}{\rm GeV}
=9.5&+0.3\left(
\frac{\alpha_s(M_Z)-0.1184}{0.0007}\right)\\
&+0.7\left(\frac{M_h}{{\rm GeV}}-125.15\right)\\
&-1.0\left(\frac{M_t}{{\rm GeV}}-173.34\right).
\label{eq:fit}
\eeq
The factor of $0.0007$ dressing $\alpha_s(M_Z)$ corresponds to the $1\sigma$ uncertainty (in 2013). The scale $\Lambda_*$ can be matched to the zero of the $\overline{\rm MS}$ quartic via $\Lambda_\lambda\simeq2\Lambda_*$ \cite{Buttazzo:2013uya}. The current LHC Higgs mass determinations are \cite{ATLAS:2023oaq,CMS:2020xrn}
\beq
M_h^{\rm ATLAS} &=125.11\pm0.11~{\rm GeV},\\
M_h^{\rm CMS}&=125.38\pm0.14~{\rm GeV},
\eeq
we will take $M_h\simeq125.2~{\rm GeV}$ as a central value. For the top quark mass we take the reconstruction result
\cite{ATLAS:2024dxp}
\beq
m_t^{\rm MC}=172.52\pm0.33~{\rm GeV}.
\eeq
The LHC measurements set the top mass parameter used in Monte Carlo event generators, rather than an unambiguous pole mass. It is common practice to take this as the $M_t^{\rm pole}$ while including an additional $0.5~{\rm GeV}$ uncertainty associated with the conversion \cite{Hoang:2020iah}. Adding the uncertainties in quadrature, we take
\beq
M_t^{\rm pole}=172.52\pm0.6~{\rm GeV}.
\eeq
For the strong coupling at the $Z$-pole we take
\cite{ParticleDataGroup:2026mpi}
\beq
\alpha_s(M_Z)=0.1180\pm0.0009.
\eeq
Giving $\log_{10}(\Lambda_*/{\rm GeV})\simeq10.18$. Hence at $1\sigma$ in the SM 
\beq
\widetilde m\equiv\Lambda_\lambda\simeq 
10^{10.5\pm0.7}~{\rm GeV}~.
\label{central}
\eeq
Thus at $2\sigma$ the quartic vanishing scale lies within the range $(1.1\times10^{9} -8.4\times10^{11})$~GeV. 

The above neglects threshold corrections due to any splitting in the spectrum around $\widetilde m$, however, we anticipate the spectrum to be tight. Since we take the Higgsino as the LSP, in Appendix \ref{ApA} we examine the threshold correction due to a small splitting of the Higgsino from $\widetilde{m}$. We find that for $m_{\widetilde{H}}>8\times10^{10}$ GeV, the Higgsino threshold corrections imply a sub-percent level change in~$\widetilde{m}$.

Having established the quartic vanishing scale, we next review the three known SUSY mechanisms that explain this phenomenon.

\subsection{Supersoft Dirac gauginos}
\label{subsec:dirac}

In the Minimal Supersymmetric Standard Model (MSSM) the quartic is inherited from the electroweak $D$-term potential
\beq
V_D&=\frac{1}{2}D^aD^a+D^a \left(g \phi^\dagger T^a\phi\right)
\eeq
and integrating out the auxiliary field $D^a$ yields
\beq
V_D=\frac{g_2^2+g_Y^2}{8}\left(|H_u^0|^2-|H_d^0|^2\right)^2.
\label{eq:Dterm}
\eeq
The light eigenstate $h=\sin\beta H_u^0+\cos\beta H_d^{0*}$ corresponds to the sole Higgs DoF that is tuned below the SUSY scale. Matching to the SM potential of eq.~(\ref{VSM}), the light Higgs quartic coupling is set by the tree-level  condition
\beq
\lambda(\widetilde m)=\frac{1}{8}\left[g_2^2(\widetilde m)+g_Y^2(\widetilde m)\right]\cos^22\beta.
\label{eq:matching}
\eeq

A problem in TeV scale SUSY with supersoft Dirac gaugino masses is that the adjoint partners responsible for the Dirac masses simultaneously suppress the MSSM $D$-term quartic, driving the Higgs quartic to zero at tree level \cite{Fox:2002bu}. It was highlighted by the present author in \cite{Unwin:2012fj} that this problem for TeV scale SUSY is a virtue for high scale SUSY as it naturally results in the vanishing of the Higgs quartic at the superpartner scale.

 Dirac masses for the gauginos can be introduced if the particle content of the MSSM is supplemented with gauge adjoint chiral superfields \cite{Fayet:1975yi,Hall:1990hq,Fox:2002bu}. For the electroweak groups, the required field content is
\beq
\mb{S}\sim({\bf 1},{\bf 1})_0,~~{\rm and}~~\mb{T}\sim({\bf 1},{\bf 3})_0.
\eeq
 Dirac masses arise from the operators of the form \cite{Fox:2002bu}
\beq
\int \dd^2\theta \frac{\sqrt{2} W'_\alpha W_i^\alpha A_i}{M_{\rm mess}}
\supset -M_{D_i} \lambda_i\psi_{A_i}+{\rm h.c.},
\eeq
where $W'_\alpha=\theta_\alpha D'$ is a hidden sector spurion, $A_i=\mb{S},\mb{T}$, and $M_{D_i}=D'/M_{\rm mess}$. The same superspace integral generates a coupling of the real adjoint scalar $\sigma^a$ to the auxiliary field $D^a$, so that for each gauge group 
\beq
V_{D_i}
=\frac{1}{2}D_i^aD_i^a
+D_i^a\left(g_i\phi^\dagger T_i^a\phi+2M_{D_i}\sigma_i^a\right)+\frac{1}{2}m_{\sigma_i}^2(\sigma_i^a)^2,
\notag
\eeq
where $m_\sigma$ denotes the additional non-supersoft mass of the real adjoint scalar. Integrating out $D^a$ and $\sigma^a$ gives the effective $D$-term potential
\beq
V_{D_i}^{\rm eff}=\frac{\epsilon_i}{2}\left(g_i\phi^\dagger T_i^a\phi\right)^2,
\eeq
with 
$\epsilon_i=\frac{m_\sigma^2}{m_\sigma^2+4M_D^2}$ where $m_\sigma$ are the adjoint scalar soft masses, and accordingly the quartic at tree level is set by
\beq
\lambda(\widetilde m)=\frac{1}{8}\left[\epsilon_2 g_2^2+\epsilon_1 g_Y^2\right]\cos^22\beta.
\label{eq:13}
\eeq
In the supersoft breaking limit $m_{\sigma_i}^2=0$,  thus $\epsilon_i=0$ and the quartic at tree level vanishes, realising $\lambda(\widetilde m)\simeq0$ \cite{Unwin:2012fj}. The physical $\sigma^a$ still acquire masses, with $m_\sigma^{\rm phys}= 2M_D$.


\subsection{Higgs sector symmetry}

An alternative to suppressing the quartic through the supersoft Dirac-gaugino mechanism above is to enforce $\lambda(\widetilde m)\simeq0$ by symmetry in the Higgs sector. From eq.~(\ref{eq:matching}), we observe that the boundary condition of eq.~(\ref{eq:boundary}) follows naturally if $\tan\beta=1$ (i.e.~$\cos^22\beta=0$). This value of $\tan\beta$ implies that at the quartic vanishing scale one has $|H_u^0|=|H_d^0|$, and inspection of eq.~(\ref{eq:Dterm}) confirms that this indeed sets the $D$-term potential to zero. There are two possible symmetries for realising the UV boundary condition (\ref{eq:boundary}),  the first is a shift symmetry \cite{Hebecker:2012qp}, whilst the second is a discrete exchange symmetry \cite{Ibanez:2012zg,Hebecker:2013lha}, both set $\tan\beta=1$ at $\widetilde{m}$, and thus $\lambda(\widetilde{m})=0$ at tree-level.

Writing the Higgs sector quadratic terms as follows
\beq
V_{\rm mass} = m_1^2|H_u|^2 + m_2^2|H_d|^2 + \left(m_3^2H_uH_d + {\rm h.c.} \right),
\eeq
where
\beq
m_1^2=m_{H_u}^2+|\mu|^2,\quad
m_2^2=m_{H_d}^2+|\mu|^2,\quad
{\rm and}\quad
m_3^2=B\mu,
\notag
\eeq 
in terms of the $\mu$ parameter coming from $W\supset \mu H_uH_d$. The simplest symmetry which sets  $|H_u^0|=|H_d^0|$ is a discrete exchange symmetry \cite{Ibanez:2012zg,Hebecker:2013lha}
\beq
H_u \leftrightarrow H_d^\dagger.
\eeq
This enforces $m_1^2=m_2^2$, then fine-tuning $|m_3^2|^2=m_1^2m_2^2$ leads to one light scalar eigenstate which is identified with the SM Higgs
\beq
H_{\rm SM} \simeq \frac1{\sqrt2} \left( H_u-H_d^\dagger \right).
\label{eq:lightstate}
\eeq
This implies $\tan\beta=1$ and at tree-level $\lambda(\widetilde m)\simeq0$.

A precursor to the exchange symmetry, which provides a particularly elegant realisatio of this idea, is provided by a shift symmetry of the K\"ahler potential
\beq
K\supset Z(X,\bar X)|H_u+H_d^\dagger|^2,
\eeq
which is invariant under 
\beq
H_u\rightarrow H_u+c, \qquad H_d^\dagger\rightarrow H_d^\dagger-c,
\eeq 
where $H_d^\dagger$ denotes $i\sigma_2H_d^*$, and $X$ denotes moduli or SUSY-breaking fields. Such structures naturally arise when the Higgs doublets descend from higher-dimensional gauge fields  \cite{Hebecker:2012qp,Hebecker:2013lha}.

After SUSY breaking this symmetry leads to
\beq
m_1^2=m_2^2=m_3^2.
\eeq
In the exact symmetry limit, the Higgs mass matrix has the massless mode of eq.~(\ref{eq:lightstate}) without additional tuning. Symmetry-breaking radiative corrections lift this mode, and obtaining the electroweak scale far below $\widetilde m$ still requires the usual high scale SUSY fine-tuning. Furthermore, the same K\"ahler structure also accommodates the Giudice-Masiero mechanism \cite{Giudice:1988yz}, which generates $\mu$ and $B\mu$ terms after supersymmetry breaking.

Note, neither the shift or discrete symmetry is exact at low energies. Yukawa couplings violate these symmetries under RG evolution, splitting  $m_{H_u}^2$ and $m_{H_d}^2$. As a result, $\lambda(\widetilde m)$ receives loop-suppressed corrections, although these are found to be small \cite{Hebecker:2012qp,Ibanez:2012zg}.

These Higgs sector symmetries are intrinsically supersymmetric and, moreover, are well motivated from string theory.
Finally, we highlight in passing that, although we focus here on SUSY realisations \cite{Hebecker:2012qp,Hebecker:2013lha,Ibanez:2012zg,Unwin:2012fj} of the UV boundary condition \eqref{eq:boundary}, non-supersymmetric routes to eq.~(\ref{eq:boundary}) also exist, notably 5D gauge-Higgs unification \cite{Gogoladze:2007qm} and Higgs Parity \cite{Hall:2018let} which is a non-SUSY (left-right chirality) symmetry of the Higgs sector.

\section{Intermediate scale Higgsino LSP}
\label{sec:Higgsino}

An attractive feature of  SUSY is that $R$-parity conservation makes the LSP a natural dark matter candidate. SUSY dark matter has been explored extensively for the TeV scale MSSM and thus it makes sense to ask the same for models of SUSY at the quartic vanishing scale $\widetilde{m}$.

\subsection{Dirac Higgsino LSP}

To ascertain whether the LSP can be a viable dark matter candidate, we should first identify what the natural candidates are for the LSP.\footnote{As an alternative, the Higgs sector symmetry models \cite{Hebecker:2012qp,Hebecker:2013lha,Ibanez:2012zg} assumed the LSP was not dark matter, but rather the QCD axion.}
In the original Dirac gaugino mechanism for setting $\lambda(\widetilde m)=0$ in \cite{Unwin:2012fj}, the LSP was taken to be a TeV scale Dirac Wino. A Dirac Wino has no tree-level vector coupling to the $Z$, and thus evades direct detection constraints. A later paper \cite{Fox:2014moa} presented a variation of  \cite{Unwin:2012fj} in which the LSP was a Weak scale Higgsino. In this case, the Higgsino tree-level cross-section must be removed by splitting the neutral Dirac state into a pseudo-Dirac pair, with a splitting $\delta$ large enough that inelastic up scattering $\chi_1 N\to\chi_2 N$ is kinematically forbidden, $\delta\gtrsim \OO(100)~{\rm keV}$, leaving only loop-induced elastic scattering. We shall discuss the inelastic case in Section \ref{sec:LZ}; first we focus on an effectively Dirac Higgsino with tree-level $Z$ exchange, and as a result, direct detection requires the Higgsino to lie at the intermediate scale or above.

The Higgsino mass is set by the $\mu$ parameter from the superpotential term $\mu H_uH_d$. In the vanilla MSSM this term is unprotected by any symmetries and naturally lies at $\mu\sim M_{\rm Pl}$, this is known as the $\mu$ problem. A natural solution to the $\mu$ problem is to forbid the operator $H_uH_d$ at tree-level and then reintroduce it via a higher dimension operator, this is the basis of the Giudice-Masiero or Kim-Nilles mechanisms \cite{Giudice:1988yz,Kim:1983dt}. In this case $\mu$ can readily be marginally (or hierarchically) lighter than the superpartner scale $\widetilde{m}$ and the Higgsino can be the LSP.

The phenomenology depends not only on the Higgsino being the LSP, but also on its mixing with the heavier electroweak gauginos. This mixing determines the size of the mass splitting between the neutral Higgsino states and thus whether the Higgsino is effectively Dirac or not. For gaugino  masses of order  $\widetilde m$, the induced neutral-Higgsino splitting is
$\delta m_0\sim  m_Z^2/\widetilde m \sim 0.1~{\rm keV}$. The LSP is thus an almost pure Higgsino, with $m_\chi\simeq|\mu|$. 

Specifically, the two neutral Weyl Higgsinos combine into the four-component field
\beq
\chi_H
=
\begin{pmatrix}
\widetilde H_u^0\\
\widetilde H_d^{0\dagger}
\end{pmatrix},
\label{eq:diracH}
\eeq
with mass $m_{\widetilde H^0}=|\mu|$.
The neutral Higgsinos are accompanied with charged partners $\widetilde H^\pm$. Electroweak radiative corrections split the charged and neutral states by \cite{Cirelli:2005uq}
\beq
\Delta m \equiv m_{\widetilde H^\pm}-m_{\widetilde H^0} \simeq350~{\rm MeV}.
\eeq
Thus the neutral Higgsino is the LSP and is stable by $R$ parity, while Higgsino number is approximately conserved in the decoupling limit considered here. Hence, the Higgsino is a potential dark matter candidate.

The hierarchy $|\mu|\lesssim\widetilde m$ can arise naturally in each of the Higgs sector constructions discussed above. In the discrete exchange-symmetry realisation \cite{Ibanez:2012zg,Hebecker:2013lha}, an ordinary $\mu$ term contributes equally to $m_1^2$ and $m_2^2$ and therefore does not spoil the condition $m_1^2=m_2^2$. In the shift-symmetric realisation \cite{Hebecker:2012qp}, the same K\"ahler structure that gives the Higgs sector symmetry  accommodates the Giudice-Masiero mechanism \cite{Giudice:1988yz} via
\beq
K \supset Z(X,\bar X)\left(H_uH_d+{\rm h.c.}\right),
\label{eq:GM}
\eeq
which simultaneously generates $\mu$ and the correlated $B\mu$ term after supersymmetry breaking.

While the shift symmetry  \cite{Hebecker:2012qp} explicitly forbids the tree-level $\mu$-term, this is not the case for either the exchange symmetry \cite{Ibanez:2012zg,Hebecker:2013lha}  or the supersoft Dirac gaugino model~\cite{Unwin:2012fj}. In the latter cases, an additional symmetry is required to solve the $\mu$ problem. Notably a PQ symmetry \cite{Peccei:1977hh} can play this role very nicely, while simultaneously addressing the Strong CP problem \cite{Kim:1983dt}. Another option is a continuous global U(1)${}_R$ symmetry (or large discrete R-symmetry $Z_R^n$) which forbids the $\mu$ term if the R-charge of the Higgsino superfields is $R[H_u]+R[H_d]\neq2$; we note that the canonical assignment is $R[H_{u,d}]=0$ \cite{Fox:2002bu,Hall:1990hq,Kribs:2007ac}.

R-symmetric SUSY \cite{Unwin:2012fj,Fox:2002bu,Hall:1990hq,Kribs:2007ac} is an elegant variation upon the traditional MSSM. That it resolves the $\mu$ problem is particularly appealing, however, it also presents a number of phenomenological hurdles. Most notably, an exact R-symmetry also forbids Majorana gaugino masses. It is anticipated that the R-symmetry will be broken in which case loop suppressed, or anomaly mediated, Majorana masses can be generated \cite{Giudice:1998xp,Randall:1998uk}.

As such, R-symmetric models present an ideal setting for extending the particle content to include gauge adjoint superpartners to give Dirac masses to the gauginos. In this manner, the vanishing of the Majorana gaugino masses is unimportant, and the gauginos can be nearly purely Dirac. Interestingly, such constructions imply an $N=1$ matter sector with an $N=2$ gauge sector and a global U(1)$_R$, which is the most symmetric scenario consistent with recovering the low-energy SM \cite{Fox:2002bu} (obtaining chirality in the SM with an $N=2$ matter sector being the principal obstruction to fully realising $N=2$ SUSY).

With the bare $\mu$ term forbidden by a discrete or continuous ($R$ or regular) symmetry, the bilinear Higgs term can then be generated in the superpotential after symmetry breaking, which we parameterise with the spurion~$X$
\beq
W
\supset
\frac{X^n}{M_{\rm mess}^{n-1}}H_uH_d,
\eeq
 leading to an effective $\mu$ term
\beq
\mu=\frac{\langle X\rangle^n}{M_{\rm mess}^{n-1}}.
\eeq
While this does not predict a definite ratio for $|\mu|/\widetilde m$, a moderate hierarchy $|\mu|\lesssim\widetilde m$ arises naturally when the symmetry breaking scale $\langle X\rangle$ lies modestly below the mediation scale $M_{\rm mess}$, with the resulting effective $\mu$ term tied to the same SUSY-breaking sector that sets $\widetilde m$.

\vspace{-2mm}
\subsection{Neutral-Higgsino splitting}
\vspace{-2mm}

As prefaced earlier, whether the Higgsino behaves as an effectively Dirac fermion for the purposes of direct detection depends on the size of the mass splitting between the neutral Higgsino states, which in turn depends on the Higgsino-gaugino mixing.
If the neutral Higgsino mixes with heavy neutral gauginos, this induces small Majorana masses and splits this state into two nearly degenerate Majorana eigenstates, sometimes called pseudo-Dirac or inelastic dark matter \cite{Tucker-Smith:2001myb} (as is the focus of Section \ref{sec:LZ}).

With reference to the models above, two specific cases emerge corresponding to whether the model respects U(1)${}_R$ or not. In the $R$-symmetric case, the Majorana gaugino masses are negligible $M_i\ll \widetilde{m}$, whereas if U(1)${}_R$ is not respected, then $M_i\sim \widetilde{m}$ (assuming a universal SUSY scale as might arise via gravity mediation). As discussed above the U(1)${}_R$ is particularly relevant to the supersoft mechanism proposed in \cite{Unwin:2012fj}.

We first consider ordinary Bino and Wino Majorana masses, $M_{1,2}\sim\widetilde m$. Integrating out the electroweak gauginos gives the neutral-Higgsino splitting \cite{Arkani-Hamed:2004ymt,Giudice:2004tc}
\beq
\delta m_0^{\rm Maj}
\simeq m_Z^2\left|\frac{s_W^2}{M_1}+\frac{c_W^2}{M_2}\right|.
\eeq
For the $R$-symmetric Dirac gaugino scenario, Majorana gaugino masses are suppressed while the electroweak gauginos acquire Dirac masses
$M_{D_i}\simeq\widetilde m$. In this case one finds (see
Appendix~\ref{ApB}) an analogous expression for the neutral-Higgsino
splitting in terms of the Dirac masses $M_{D_i}$ and the adjoint fermion
Majorana masses $M_{A_i}$,
\beq
\delta m_0^{\rm Dirac}
\simeq
m_Z^2
\left|s_W^2
\frac{M_{A_1}-\mu\sin2\beta}{M_{D_1}^2}+c_W^2
\frac{M_{A_2}-\mu\sin2\beta}{M_{D_2}^2}
\right|.
\eeq
In the supersoft case we take $|M_{A_i}|\ll M_{D_i}\sim\widetilde m$, as expected for an approximately conserved $R$ symmetry. Provided the expansion condition in Appendix~\ref{ApB} holds, the resulting splitting is typically smaller than in the ordinary Majorana case, for which $M_{1,2}\sim\widetilde m$ gives
\beq
\delta m_0^{\rm Maj}
\sim \frac{m_Z^2}{\widetilde m}
\sim 0.1~{\rm keV}
\left(\frac{10^{11}~{\rm GeV}}{\widetilde m}\right).
\eeq
This is well below the $\OO(100~{\rm keV})$ inelastic threshold of direct detection experiments. Thus, the neutral Higgsino retains the full effective Dirac $Z$-mediated scattering rate.

Finally, we note that Higgsino-number-violating higher-dimensional operators can also split the neutral state. For example, the dimension five operator
\beq
\mathcal L_{\Delta H=2}
\supset \frac{1}{\Lambda_5} \left(H^\dagger\widetilde H\right)^2,
\eeq
which implies a characteristic splitting $\delta m_0 \sim v^2/\Lambda_5$.
Requiring the splitting to remain below the characteristic inelastic
threshold gives
\beq
\Lambda_5
\gtrsim
\frac{v^2}{100~{\rm keV}} \simeq 6\times10^8~{\rm GeV}.
\eeq
Since we have implicitly assumed that the theory is SM below the LSP Higgsino mass $\mu$, this implies that generically $\Lambda_5\gtrsim\mu\sim 10^{11}$ GeV. It follows that higher dimension operators also give splittings below the direct-detection inelastic threshold. Thus, provided that the gauginos have masses of order $\widetilde{m}$, the neutral Higgsino will behave in direct detection experiments as an effectively Dirac electroweak doublet. (The inelastic scenario of Section \ref{sec:LZ} instead assumes an alternative hierarchy with $\mu\ll M_{1,2}\ll\widetilde{m}$).

\section{Direct detection}
\label{sec:DD}

For a neutral Higgsino expressed as the Dirac fermion of eq.~(\ref{eq:diracH}) there is a pure vector interaction with the~$Z$
\beq
\mathcal L_Z = \frac{g_2}{2c_W} Z_\mu\overline{\chi}_H\gamma^\mu\chi_H .
\eeq
The coupling is non-vanishing because $\chi_H$ is effectively Dirac. In contrast, for a pair of Majorana states, the corresponding diagonal vector currents vanish, and the $Z$ interaction is off-diagonal. This does not by itself suppress the scattering rate, the interaction is suppressed only when the neutral-state splitting makes the inelastic transition kinematically inaccessible, see e.g.~\cite{Tucker-Smith:2001myb,Fox:2014moa,Bernal:2026clv}.

For direct detection experiments, the momentum transfer satisfies $q^2\ll m_Z^2$, and one can integrate out the $Z$ leading to the effective operator
\beq
\mathcal L_{\rm eff} = \sqrt{2}G_F  \overline{\chi}_H\gamma_\mu\chi_H \sum_q \overline q\gamma^\mu \left(g_V^q-g_A^q\gamma^5\right)q ,
\eeq
where 
\beq
g_V^q=T_3^q-2Q_qs_W^2.
\eeq
For the Higgsino $T_3(\widetilde H_u^0)=-\tfrac12$ and  $T_3(\widetilde H_d^0)=+\tfrac12$.
Only the vector quark current contributes coherently to spin-independent (SI) scattering. Moreover, one can express the effective couplings at the nucleon level in the following manner
\beq
f_p &= 2g_V^u+g_V^d = \frac12-2s_W^2\simeq0.038,\\[10pt]
f_n&= g_V^u+2g_V^d = -\frac12 ,
\eeq
where we have taken $s_W^2\approx0.231$. Since $f_p \ll f_n$ the coherent scattering amplitude is dominated by the neutron contribution.

The SI cross section for scattering from a nucleus with atomic number $Z$ and mass number $A$ is
\beq
\sigma_A^{\rm SI} = \frac{2G_F^2\mu_{\chi A}^2}{\pi} \left[ Zf_p+(A-Z)f_n \right]^2 ,
\eeq
where $\mu_{\chi A}$ is the dark matter-nucleus reduced mass.  The corresponding
nucleon level cross sections are
\beq
\sigma_Z^n &=\frac{G_F^2\mu_{\chi n}^2}{2\pi}
\simeq 7.4\times10^{-39}~{\rm cm}^2 ,\\[10pt]
\sigma_Z^p &= \frac{G_F^2\mu_{\chi p}^2}{2\pi} \left(1-4s_W^2\right)^2
\simeq 4.3\times10^{-41}~{\rm cm}^2 .
\label{eq:sigmap}
\eeq
Here $\mu_{\chi p}$ is the dark matter-nucleon reduced mass (similarly $\mu_{\chi n}$). For the heavy dark matter under consideration $m_\chi\gg m_p, m_n$ and thus $\mu_{\chi p}\approx m_p$ and $\mu_{\chi n}\approx m_n$.  In particular, in this heavy dark matter limit, note that the scattering cross sections become independent of $m_\chi$.

LZ presents the leading limit on SI dark matter scattering  \cite{LZ:2024zvo}. The LZ collaboration typically reports their limits in terms of per-nucleon cross section under the isospin-conserving assumption, i.e.~that the proton and neutron couplings are equal.  Since this does not hold in our case (cf.~eq.~\eqref{eq:sigmap}) we need to rescale our cross sections to obtain an appropriate recasting of the limits. Specifically, for a xenon isotope $i$ with mass number $A_i$, neutron number $N_i=A_i-Z$, and natural abundance $\eta_i$, the conversion factor is \cite{Feng:2011vu}
\beq
\sigma_{\rm eff}^{\rm Xe} = \mathcal R_{\rm Xe} \sigma_Z^n ,
\label{eq:sigmaeff}
\eeq
where
\beq
\mathcal R_{\rm Xe} \equiv \frac{\sum_i\eta_i\mu_{\chi A_i}^2\left[Z(f_p/f_n)+N_i\right]^2}{\sum_i\eta_i\mu_{\chi A_i}^2A_i^2} \simeq0.311 ,
\eeq
summing over the naturally occurring xenon isotopes.

\begin{figure}[t]
\centering
\hspace*{-5mm}
\includegraphics[width=1.1\columnwidth]{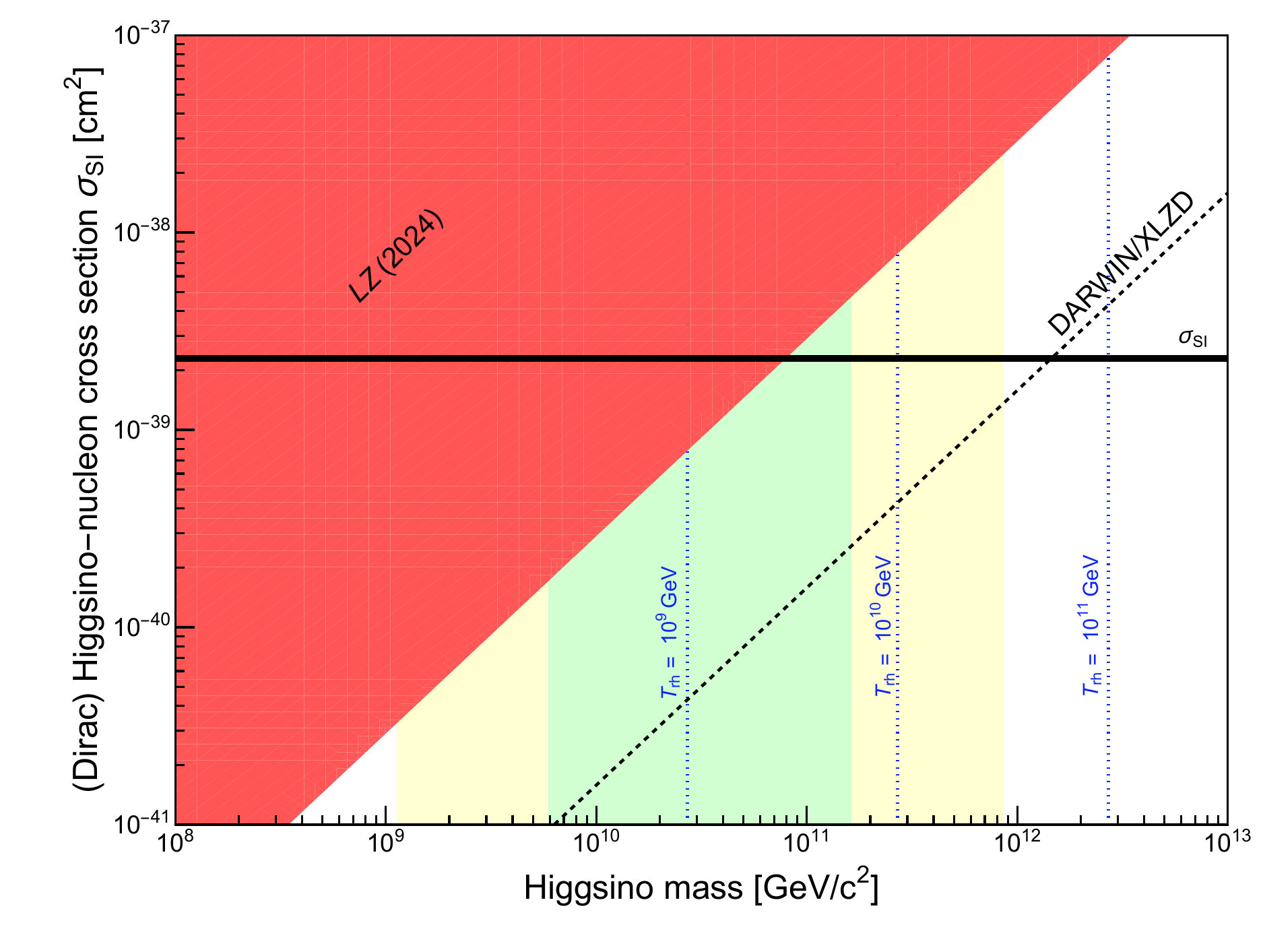}
\caption{
Spin-independent direct detection of the effectively Dirac Higgsino. The horizontal black line shows the predicted xenon-equivalent cross section, compared with the LZ 90\% C.L.\ exclusion limit \cite{LZ:2024zvo} (solid black), its expected $68\%$ and $95\%$ sensitivity bands (green and yellow), and the projected DARWIN/XLZD sensitivity \cite{Baudis:2024jnk} (black dashed). The blue dotted lines denote representative reheating temperatures, while the red vertical band indicates the scale $\widetilde m$ at which the SM Higgs quartic coupling vanishes, including the propagated uncertainties in $(M_t,\alpha_s,M_h)$.}
\label{fig:DD}
\end{figure}

It follows that the effective scattering cross section, relevant for comparison to the LZ limits is\footnote{The SI cross sections given in
\cite{Bernal:2026clv} are too small by a factor of 16, corresponding to a factor of two in each of the dark matter and nucleon $Z$ couplings in $\sigma_{\rm SI}$. The error is restricted to the evaluation of $\sigma_{\rm SI}$, and the electroweak couplings reported ($g_{V,A}^{u,d}$, etc.) are correct.  The freeze-in calculations are unaffected. That analysis compared the neutron-level cross section directly against the LZ isospin-conserving limit, without the rescaling of eq.~(\ref{eq:sigmaeff}). Including both effects  the limit strengthens by $16\mathcal R_{\rm Xe}\simeq5$. The numerical inputs in Eqs.~(22) \& (23) of \cite{Bernal:2026clv} imply an uncorrected bound of $1.6\times10^{10}~{\rm GeV}$, rather than the quoted $1.4\times10^{10}~{\rm GeV}$, applying both corrections gives $m_\chi\gtrsim7.8\times10^{10}~{\rm GeV}$.}
\beq
\sigma_{\rm eff}^{\rm Xe}
\simeq 0.311 \sigma_Z^n
\simeq 2.3\times10^{-39}~{\rm cm}^2 .
\eeq
At masses well above the target nucleus, the recoil spectrum is largely insensitive to increasing $m_\chi$, and the local dark matter mass density remains approximately fixed. The corresponding number density therefore scales as $n_\chi=\rho_\chi/m_\chi$, implying a dark matter flux $\Phi_\chi\propto1/m_\chi$. Since the event rate is proportional to the incident flux, $R\propto\sigma_A^{\rm SI}/m_\chi$, the SI exclusion limit weakens linearly with the dark matter mass for $m_\chi\gg m_Z$. We obtain the high-mass limits shown in Fig.~\ref{fig:DD} by linearly extrapolating the LZ 4.2 tonne-year result \cite{LZ:2024zvo} to the high-mass regime.
For the mass independent cross section of eq.~(\ref{eq:sigmaeff}) this implies a lower mass bound on the Dirac Higgsino:
\beq
m_\chi \gtrsim 7.8\times10^{10}~{\rm GeV}.
\eeq

Notably, the LZ bound intersects the upper part of the range selected by the vanishing SM Higgs quartic. The two requirements together
select
\beq
m_\chi \simeq |\mu| \simeq 10^{11.4\pm0.5}~{\rm GeV}~.
\eeq
Moreover, the direct detection constraint and the requirement that the Higgsino is the LSP imply the following tight window on the Higgsino mass parameter
\beq
7.8\times10^{10}~{\rm GeV}\lesssim |\mu|\lesssim\widetilde m.
\eeq
The hierarchy is therefore compressed to $ \widetilde{m}/m_\chi \lesssim 10$. In a Giudice-Masiero realisation with 
\beq
|\mu|=\langle X\rangle^n/M_{\rm mess}^{n-1}\equiv c_\mu\widetilde m,
\eeq this corresponds to a coefficient $c_\mu \sim 0.1-1$.

Figure~\ref{fig:DD} displays the Higgsino scattering cross section in the isospin-conserving convention used by LZ.  The intersection of the scattering cross section and vanishing quartic region,  lies partially within the present exclusion region, and further exposure will continue to constrain the parameter region consistent with the vanishing quartic boundary condition. Moreover, all of this parameter space lies above the neutrino fog, implying that this scenario will be decisively tested at future direct detection experiments such as DARWIN/XLZD \cite{Baudis:2024jnk}.

While direct detection provides a decisive probe of this scenario, other experimental approaches are anticipated to be less successful. Collider production is implausible at an intermediate mass scale near $10^{11}$~GeV. One could search for Higgsino dark matter annihilation into electroweak final states ($\chi_H\overline{\chi}_H\rightarrow W^+W^-, ZZ,Zh$), with associated high-energy photon and neutrino emission. However, the diffuse indirect detection constraints remain many orders of magnitude weaker than
the direct detection requirement throughout the relevant mass range, as discussed in \cite{Bernal:2026clv}.
Similarly, the scattering cross sections are too small for signatures in atmospheric events or geological samples, e.g.~\cite{Ebadi:2021cte,SinghSidhu:2019cpq}.

\section{Relic abundance from Boltzmann suppressed freeze-in}
\label{sec:BSFI}

Dark matter with $m_\chi\sim10^{11}$~GeV cannot be a thermal relic due to freeze-out via electroweak interactions. Thus for the Higgsino LSP to be the dark matter its relic density must be set by a different mechanism. An alternative prospect is dark matter freeze-in \cite{Hall:2009bx,Elahi:2014fsa} in which the Higgsino abundance is initially negligible and the relic abundance is set via production processes initiated within the SM thermal bath. The issue with traditional dark matter freeze-in is that the electroweak boson mediated production rate is too high and the Higgsino will generically thermalise, subsequently freeze out, and ultimately fail to reproduce the observed relic abundance. 
A viable approach to obtain the observed dark matter relic density is found within Boltzmann suppressed freeze-in \cite{Giudice:2000ex,Cosme:2023xpa,Bernal:2025fcl} (A.K.A.~`freeze-in at stronger coupling'). 

Specifically, if the Higgsino mass exceeds the reheat temperature of the Universe $T_{\rm rh}$, then the Higgsino production rate is exponentially suppressed and as a result it never thermalises and the correct dark matter abundance can be realised for an appropriate choice of $m_\chi/T_{\rm rh}>1$.
The hierarchy of scales is thus
\beq
m_Z,m_h,m_t \ll T_{\rm rh} < |\mu| < \widetilde m.
\eeq

Since $T_{\rm rh}$ is well above the electroweak scale, all SM particles are effectively massless and remain in thermal equilibrium throughout Higgsino production. In contrast, for the remaining superpartners $T_{\rm rh}<\widetilde m$,  their thermal abundances are exponentially suppressed $\exp(-\widetilde m/T_{\rm rh})$. As $\widetilde m>|\mu|$, this suppression is parametrically stronger than that governing Higgsino production. Thus the heavier superpartners are never appreciably produced and do not impact the relic density.

The freeze-in dynamics hence can be described by an effective field theory (EFT) consisting of the SM supplemented by the vector-like Higgsino doublets. Since the neutral and charged Higgsinos are only 350 MeV apart, both DoFs are retained in the EFT.  We highlight that, since the Higgsinos form a vector-like pair of electroweak doublets, this is a SUSY realisation of the ``minimal freeze-in dark matter'' scenario presented in~\cite{Bernal:2026clv}. The neutral Higgsinos are the LSP, and the charged Higgsinos decay via $\widetilde{H}^\pm\rightarrow\widetilde{H}^0\pi^\pm$. Notably, freeze-in and decay of $\widetilde{H}^\pm $ should be tracked since it perturbs the results by an $\mathcal{O}(1)$ factor.

Integrating out the heavy superpartner spectrum induces higher-dimensional interactions between Higgsinos and SM fields. However, these operators provide only subleading corrections to the renormalisable electroweak gauge interactions already present in the EFT. Accordingly, Higgsino production is dominated by the tree-level electroweak processes mediated by off-shell $Z$, $\gamma$, and $W^\pm$ bosons, while the higher-dimensional operators can be safely neglected.
We also note that since $m_{3/2}\sim\widetilde m$ and thus $T_{\rm rh}<m_{3/2}$ the usual gravitino problem is evaded.

The freeze-in abundances of $\widetilde H^0$ and $\widetilde H^\pm$ can be calculated in terms of the total yield $Y_\chi=n_\chi/s$, where $n_\chi$ counts the full Higgsino multiplet. Assuming radiation domination after instantaneous reheating at temperature $T_{\rm rh}$ and $x_{\rm rh}\equiv m_\chi/T_{\rm rh}\gg1$, the Higgsino yield is
\beq
Y_{\rm FI}\simeq
\frac{135\sqrt{10}\,g_H^2}{64\pi^6}
\frac{a_H}{\gss\sqrt{\gs}}
\frac{M_{\rm Pl}}{m_\chi}
(2x_{\rm rh}+1)e^{-2x_{\rm rh}},
\label{eq:yield}
\eeq
where $g_H=8$ is the internal DoF of the Higgsino multiplet, and $a_H$ is given by \cite{Arkani-Hamed:2006wnf}
\beq
a_H=\frac{21g_2^4+3g_2^2g_Y^2+11g_Y^4}{512\pi}.
\eeq
The effective annihilation rate is $\langle\sigma_{\rm eff}v\rangle\simeq a_H/m_\chi^2$ which includes annihilation and coannihilation of the charged and neutral states. Here $\gs$ is the relativistic energy DoF, $\gss$ is the effective entropy DoF, and $M_{\rm Pl}$ is the reduced Planck mass. The charged Higgsinos subsequently decay into the neutral states, preserving the total yield. Note the explicit Boltzmann suppression factor in eq.~(\ref{eq:yield}).

It remains to identify the parameter values for which the correct abundance is obtained. 
From the yield of eq.~(\ref{eq:yield}), we obtain the relic density
\beq
\Omega_\chi h^2\simeq
0.12\left(
\frac{m_\chi Y_{\rm FI}}
     {4.38\times10^{-10}~{\rm GeV}}
\right).
\eeq
When the instantaneous reheating assumption holds, there are only two free parameters: the reheat temperature $T_{\rm rh}$ and the Higgsino mass $m_\chi=|\mu|$.
Matching the observed abundance $\Omega_{\rm DM}h^2\simeq0.12$ \cite{Planck:2018vyg}, with $g_\star=g_{\star s}=106.75$, gives $x_{\rm rh}\simeq27$.

In Figure~\ref{fig:DD} we overlay the direct detection constraints with representative reheating temperatures (assuming instantaneous inflaton decay). This indicates the reheating temperature required for a Higgsino LSP of a given mass to reproduce the observed dark matter relic abundance via Boltzmann suppressed freeze-in.

Notably, $\Omega_{\rm DM}$ is exponentially sensitive to changes to $x_{\rm rh}$. The above assumes instantaneous reheating, if the reheating process is non-instantaneous the SM thermal bath can have a more complicated evolution, reaching a maximum temperature $T_{\rm max}>T_{\rm rh}$ \cite{Giudice:2000ex}. The production between $T_{\rm max}$ and $T_{\rm rh}$ can impact the required value of $T_{\rm rh}$ and  $T_{\rm max}$ \cite{Bernal:2025fcl,Bernal:2026vbg,Bernal:2019mhf}. Since the abundance is exponentially sensitive, modest changes in parameter values will result, but it does not alter the overall picture.

Finally, we note that for $m_\chi\simeq27T_{\rm rh}$, as needed to reproduce $\Omega_{\rm DM}$, the Higgsinos do not thermalise with the SM bath. This can be seen by comparing the production rate to Hubble, and requiring that $\Gamma_\chi=n_{\chi,{\rm eq}}\langle\sigma_{\rm eff}v\rangle<H$ for all temperatures, but most stringently at $T_{\rm rh}$. Evaluating for $m_\chi\sim10^{11}$~GeV and $x_{\rm rh}\simeq27$ we find
\beq
\frac{\Gamma_\chi}{H}\Big|_{T_{\rm rh}}
\simeq
\frac{g_Ha_H}{(2\pi)^{3/2}}
\sqrt{\frac{90}{\pi^2g_\star}}
\frac{M_{\rm Pl}}{m_\chi}
x_{\rm rh}^{1/2}e^{-x_{\rm rh}}
\ll1.
\eeq
Thus, the same exponential suppression that provides the desired dark matter abundance also automatically prevents thermalisation.

\section{PeV Inelastic Higgsino Dark Matter and the LZ High Recoil Event}
\label{sec:LZ}

While this work has largely focused on effectively Dirac Higgsino dark matter which scatters elastically, a small Majorana splitting of the neutral Higgsinos can instead lead to inelastic scattering. Such inelastic Higgsino dark matter has recently received renewed attention following the tentative high-recoil event observed at LZ \cite{LZ:2026axp}.\footnote{We note that Higgsino dark matter is not the only particle physics explanation, alternatives include more general models of inelastic dark matter e.g.~\cite{Lee:2026xxh,Smirnov:2026aqk,Su:2026rwz}, certain elastic scattering operators \cite{LZ:2026axp,Unwin:2026rdp,Alhazmi:2026efz}, or boosted dark matter
\cite{Alhazmi:2026efz,Heikinheimo:2026kwp}.} While initial interest focused on the thermal Higgsino case  \cite{Freese:2026sga,Wu:2026nhi,Fan:2026kxx}, it was quickly realised that the empty high-energy sideband \cite{Rodd:2026tyn} and solar-capture constraints \cite{Pospelov:2026ewn} exclude this possibility. An interesting alternative highlighted by Langhoff \cite{Langhoff:2026ujr} focused on non-thermal Higgsinos with masses $m_\chi\sim10^5$-$10^6\,\mathrm{GeV}$ and neutral-state splittings $\delta\sim330$-$480\,\mathrm{keV}$ which evades the sideband and solar constraints. It was shown in \cite{Langhoff:2026ujr} that an appropriate Higgsino relic density can be obtained if freeze-out occurs during a period of early matter domination followed by entropy dilution. We highlight Boltzmann suppressed freeze-in via electroweak gauge bosons as an alternative route to the correct relic density. It is quite conceivable that the scalar superpartners and heavy Higgs states lie near $\widetilde m\sim10^9~{\rm GeV}$, close to the scale at which the Higgs quartic vanishes, while the Higgsinos and electroweak gauginos are lighter by a moderate hierarchy.

Similar to earlier sections, the Higgsino remains the LSP with $m_\chi\simeq|\mu|$,  the important difference is that we now take the Majorana bino and wino masses to be of order $M_{1,2}=\mathcal O(10^7)\,\mathrm{GeV}$. The remaining superpartners and heavy Higgs states are placed at $\widetilde{m}\sim10^9$ GeV. For $|\mu|\ll|M_{1,2}|$, the neutral-Higgsino splitting is~\cite{Nagata:2014wma}
\begin{equation}
 \delta\simeq m_Z^2\left|
 \frac{s_W^2}{M_1}+\frac{c_W^2}{M_2}\right|.
\end{equation}
For $M_1=M_2$, the splitting $\delta\simeq330$-$480~{\rm keV}$  preferred by LZ corresponds at leading order to  \cite{Langhoff:2026ujr} 
\beq
M_1\simeq M_2\simeq2\times10^7~{\rm GeV}.
\eeq
For this split spectrum we assume that the heavy Higgs sector remains near the SUSY scale $\widetilde{m}$. For $\tan\beta\simeq1$ and $m_A\sim\widetilde m$, this requires
$B\mu\sim\widetilde m^2/2$ \cite{Arkani-Hamed:2006wnf}, so that $B\mu$ cannot in general be tied simply to $\mu\widetilde m$ when $\mu\ll\widetilde m$. Thus we treat $B\mu$ as an independent soft parameter. 

\begin{figure}[t]
 \centering
  \includegraphics[width=0.95\columnwidth]{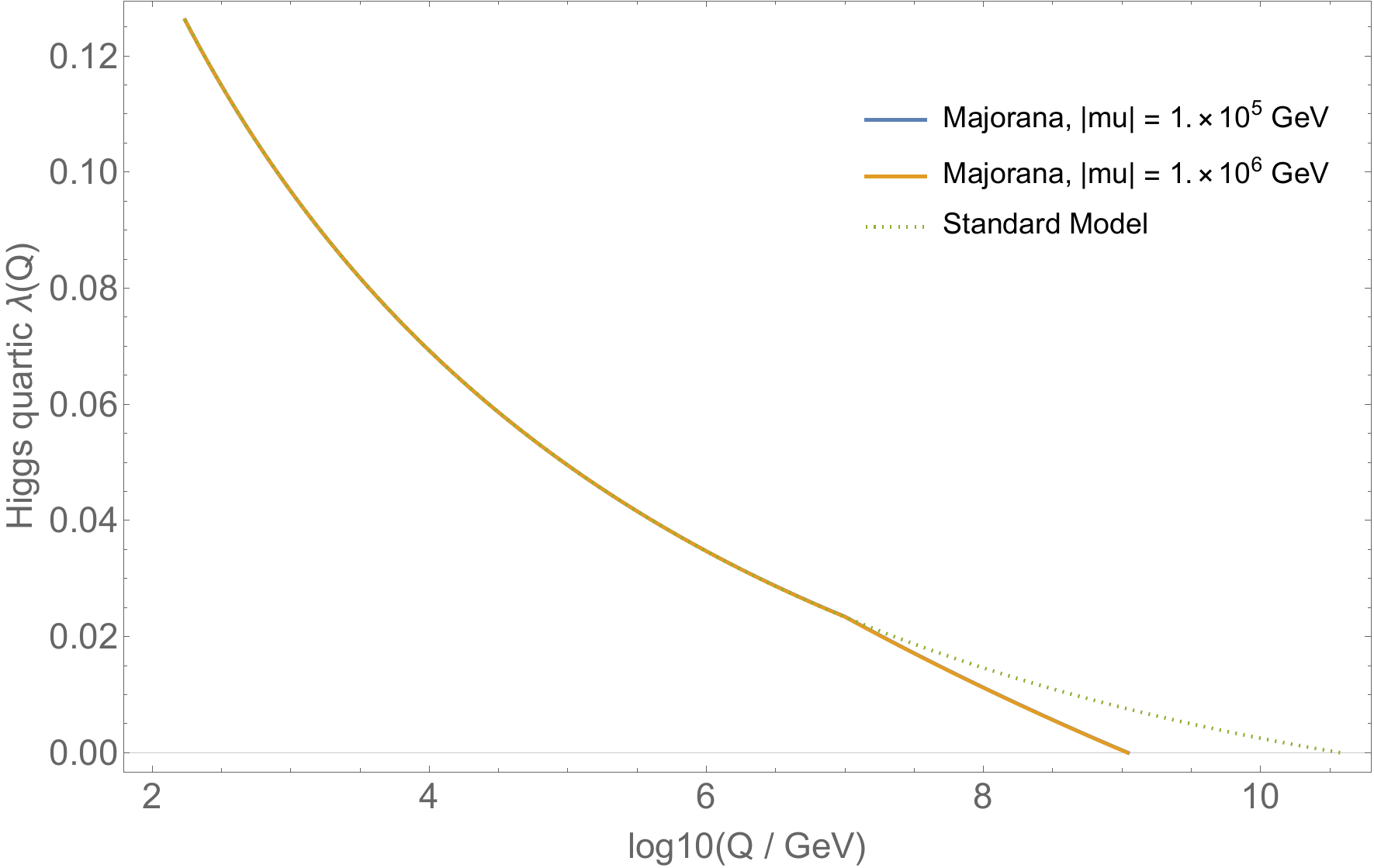}
 \vspace{-2mm} \caption{Higgs-quartic evolution for $|\mu|=10^5,~10^6\,\mathrm{GeV}$ with electroweak gauginos at
 $10^7\,\mathrm{GeV}$, compared with SM running.  In both cases, we show only the central value. \label{fig:run}}
 \vspace{-4mm}
\end{figure}

Notably, lowering the Higgsinos and electroweak gauginos below $\widetilde m$ changes the running between their respective mass thresholds and the scalar scale. Below $|\mu|$ the running is that of the SM, while above $|\mu|$ the Higgsino doublets modify the electroweak gauge-coupling $\beta$ functions. With the gauginos still integrated out, the Higgsinos
have no renormalisable coupling to the light Higgs, so their effect on the one-loop quartic running is indirect. Above $M_{1,2}$, however, the Higgs-Higgsino-gaugino interactions also contribute directly to the quartic $\beta$ function. We evolve these interactions together with the SM gauge and Yukawa couplings, imposing their supersymmetric boundary conditions at $\widetilde m$, as shown in Figure~\ref{fig:run}.
We find that for $|\mu|=10^5-10^6~{\rm GeV}$ and $M_{1,2}=10^7~{\rm GeV}$ the quartic vanishes near $\widetilde m\sim10^9~{\rm GeV}$. Thus this variant retains the connection between the scalar mass scale $\widetilde m$ and the vanishing Higgs quartic scale.

We assume instantaneous reheating into a thermal SM bath at $T_{\rm rh}$, a negligible initial Higgsino abundance, and radiation domination with no subsequent entropy release. The inelastic splitting is irrelevant during production, both because $\delta/T_{\rm rh}\ll1$ and because electroweak symmetry is restored. Thus for $T_{\rm rh}<m_\chi$, production is Boltzmann suppressed and proceeds as in Section \ref{sec:BSFI}, with the abundance $Y_{\rm FI}$ set by eq.~(\ref{eq:yield}). Since the explicit mass dependence cancels in $m_\chi Y_{\rm FI}$, the observed abundance again requires $m_\chi/T_{\rm rh}\simeq27$. For Higgsinos with the appropriate mass to explain the LZ event this implies
\begin{equation}
 T_{\rm rh}\simeq\frac{m_\chi}{27}
 \simeq
  3.7\times \begin{cases}
10^3~\mathrm{GeV}, & m_\chi=10^5~\mathrm{GeV},\\
10^4~\mathrm{GeV}, & m_\chi=10^6~\mathrm{GeV}.
 \end{cases}
\end{equation}
At these temperatures the Higgsino production rate remains below the Hubble rate, so that the produced population never reaches thermal equilibrium and annihilation of the produced population is negligible.

Such a hierarchical spectrum with $|\mu|\sim10^5$-$10^6~{\rm GeV}$ and $M_{1,2}\sim10^7~{\rm GeV}$, while the scalar superpartner lie at $\widetilde m\sim10^9~{\rm GeV}$ is quite typical for high-scale SUSY \cite{Hall:2009nd,Arkani-Hamed:2004ymt,Giudice:2004tc}. The Majorana gaugino masses can be loop-suppressed relative to the scalars, and the Higgsino mass is set by the Giudice-Masiero mechanism \cite{Giudice:1988yz}. 
Notably, taking $\widetilde m\sim10^9~{\rm GeV}$ aligns the scalar superpartner scale with the central value of the modified quartic-vanishing scale (cf.~Fig.~\ref{fig:run}).
Thus not only does Boltzmann suppressed freeze-in provide an alternative cosmological origin for the heavy inelastic Higgsino interpretation of the LZ high-recoil event, it also retains this possible connection to a vanishing quartic at a lower SUSY threshold.

\section{Concluding remarks}
\label{sec:conc}

The energy at which the Higgs quartic vanishes could be indicative of the scale of physics beyond the SM. Building on earlier work which established the SUSY UV boundary conditions \cite{Hebecker:2012qp,Hebecker:2013lha,Ibanez:2012zg,Unwin:2012fj} that realise $\lambda(\widetilde m)=0$, we have shown that these frameworks naturally accommodate Higgsino dark matter with a mass near the quartic vanishing scale, providing a motivated heavy dark matter candidate. In the heavy gaugino regime considered in Sections \ref{sec:Higgsino} \& \ref{sec:DD},  the Higgsino has an unsuppressed coupling to the $Z$, direct-detection constraints require $m_\chi>7.8\times10^{10}~{\rm GeV}$ (cf.~Fig.~\ref{fig:DD}). Such a heavy Higgsino LSP remains consistent with the quartic vanishing scale, but implies a narrow experimentally motivated mass window, $m_\chi<\widetilde m \lesssim8.4\times10^{11}~{\rm GeV}$. Accordingly, this scenario is already being constrained by LZ and can be decisively tested at DARWIN/XLZD \cite{Baudis:2024jnk}. 
We also highlighted in Section \ref{sec:LZ} that lighter gauginos ($|\mu|\ll M_{1,2} \ll \widetilde{m}$) permit an inelastic realisation. Lowering the Majorana electroweak gauginos splits the neutral Higgsinos and allows PeV inelastic Higgsino dark matter, as motivated by the tentative LZ high-recoil event \cite{LZ:2026axp}. In both cases Boltzmann suppressed freeze-in can reproduce the observed abundance while the scalar superpartners remain near the quartic vanishing scale.

A second independent prediction of this framework comes from improved determinations of the low-energy SM parameters. Since the quartic vanishing scale is highly sensitive to the measured values of $M_t$ and $\alpha_s$, future reductions in their uncertainties will provide an independent test of the intermediate-scale prediction, complementary to direct detection, in much the same way that precision gauge coupling measurements test grand unification. The framework favours a downward shift in the measured top-quark mass and/or an upward shift in the strong coupling relative to their present central values. For the Dirac case compatibility with the direct detection floor requires $M_t\simeq172.1$~GeV or $\alpha_s(M_Z)\simeq0.1190$, within roughly $1\sigma$ from the current determinations.

We also highlight that intermediate scale SUSY remains compatible with gauge coupling unification. With only the MSSM field content (as in the case of Higgs sector symmetry models), placing the superpartner spectrum at $\widetilde m\sim10^{11}$ GeV leads to unification which is comparable to that of the SM \cite{Hall:2009nd}. Precision unification then depends sensitively on GUT scale threshold corrections. 
For the supersoft Dirac gaugino model, gauge coupling running is modified by the adjoint chiral superfields above $\widetilde{m}$ (relative to the MSSM). The quality of unification is somewhat degraded, however gauge coupling unification in this class of models can still be realised in settings with large calculable threshold corrections, such as F-theory GUTs \cite{Davies:2012vu,Hebecker:2014uaa}.
While the UV boundary condition \eqref{eq:boundary} does not eliminate the fine-tuning associated with the electroweak scale, it relates the observed Higgs boson mass to a predictive boundary condition on the Higgs quartic coupling.

The framework presented here is motivated by the apparent absence of new physics at the LHC, the lack of deviations in precision measurements (such as flavour and electroweak precision tests), and the prospect that the vanishing of the Higgs quartic coupling reflects a fundamental ultraviolet boundary condition. All of these point toward new physics only at the intermediate scale, or beyond. Accordingly, this scenario is not only theoretically well motivated but is also experimentally informed.  Moreover, in the effectively Dirac realisation, the Higgs quartic vanishing scale, the dark matter relic abundance, and direct detection bounds define a remarkably predictive framework, which is falsifiable within the next generation of direct detection experiments.

\vspace{6mm}\noindent {\bf Acknowledgments.}
This work was supported by NSF grant PHY-2209998.
J.U. is grateful to New College, Queen's College, and the Rudolf Peierls Centre for their hospitality.
Claude/Codex aided in the analysis.

\appendix


\vspace{3mm}
\section{Higgsino correction to the quartic vanishing scale}
\label{ApA}

The estimate of eq.~(\ref{central}) assumes SM running between $m_t$ and $\widetilde m$. In the spectrum of interest, however, the Higgsino lies an $\mathcal{O}(1-10)$ factor below the remaining superpartners. Thus, above the scale $m_\chi=|\mu|$ the evolution of $\lambda$ will be slightly deflected relative to SM running. Here we compute the resulting shift in the scale $\Lambda_\lambda$ at which $\lambda$ vanishes, treating $m_\chi$ as a free parameter on the same footing as $M_t$ in eq.~(\ref{eq:fit}).
For the Dirac Higgsino case, below $\widetilde m$ the EFT consists of the SM supplemented by the two Higgsino doublets. Once the electroweak gauginos have been integrated out, the Higgsino has no renormalisable coupling to the Higgs. Thus at one-loop the Higgsino threshold is trivial
\beq
\lambda^{\rm SM+\widetilde H}(m_\chi) = \lambda^{\rm SM}(m_\chi).
\eeq
The leading Higgsino correction instead arises from the one-loop Higgsino contribution to the electroweak gauge-coupling $\beta$-functions inserted into the one-loop running of $\lambda$. Consequently, the first non-vanishing correction appears at two-loop order and is parametrically \cite{Giudice:2011cg}
\beq
\Delta\lambda(Q) \sim \sum_{i=1}^2 \frac{\Delta b_i g_i^6}{(16\pi^2)^2} \ln^2 \left(\frac{Q}{m_\chi}\right).
\eeq
For the two Higgsino Weyl doublets $\Delta b_Y=\Delta b_2=\frac{2}{3}$. 

To clarify the discussion we define quartic vanishing obtained using SM running only as $\Lambda_\lambda^{(0)}$ and we write the corrected vanishing scale as $\Lambda_\lambda$ (reserving $\widetilde{m}$ for the superpartner scale).
Then, evaluating the electroweak couplings near the SM crossing gives
\beq
\Delta\lambda(\Lambda_\lambda^{(0)}) &\simeq 2.5\times10^{-6} \ln^2 \left(\frac{\Lambda_\lambda^{(0)}}{m_\chi}\right).
\label{eq:del0}
\eeq

In the parameter space relevant to this scenario, direct detection requires $m_\chi\gtrsim7.8\times10^{10} {\rm GeV}$, while the quartic boundary
condition gives $\widetilde m\simeq\Lambda_\lambda\lesssim8.4\times10^{11} {\rm GeV}$. Since $m_\chi<\widetilde m$  the maximum hierarchy is restricted to be 
$\widetilde m/m_\chi \lesssim10$. Normalising to the maximum separation, the correction to the coupling is
\beq
\Delta\lambda(\Lambda_\lambda^{(0)})\simeq 1.4\times10^{-5}
\left(\frac{\ln(\Lambda_\lambda^{(0)}/m_\chi)}{\ln 10.6}\right)^2.
\label{eq:del}
\eeq

The correction is small but positive, and thus the Higgsinos shift the scale at which the quartic vanishes to slightly higher energy.
We next identify the magnitude of the shift in the vanishing scale $\Delta\Lambda_\lambda$, notably the corrected crossing scale satisfies
\begin{tightequation}
\lambda^{\rm SM}(\Lambda_\lambda)+\Delta\lambda(\Lambda_\lambda)=0.
\end{tightequation}
Expanding both terms about $\Lambda_\lambda^{(0)}$ gives
\beq
0
&\simeq
\lambda^{\rm SM}(\Lambda_\lambda^{(0)}) +\beta_\lambda^{(0)}\ln \left(\frac{\Lambda_\lambda}{\Lambda_\lambda^{(0)}}\right) +\Delta\lambda(\Lambda_\lambda^{(0)}).
\eeq
We then use that for a small change in scale 
\begin{tightequation}
\ln\left(\Lambda_\lambda/\Lambda_\lambda^{(0)} \right)
\simeq \Delta\Lambda_\lambda/\Lambda_\lambda^{(0)}.
\end{tightequation}
Moreover, by definition $\lambda^{\rm SM}(\Lambda_\lambda^{(0)})=0$, thus it follows that the fractional shift in the vanishing scale is 
\begin{tightequation}
\frac{\Delta\Lambda_\lambda}{\Lambda_\lambda^{(0)}}
\simeq \frac{\Delta\lambda(\Lambda_\lambda^{(0)})} {|\beta_\lambda^{(0)}|}.
\label{eq:shift}
\end{tightequation}
The fractional shift is therefore determined by two quantities, the slope of the SM quartic at that scale $|\beta_\lambda^{(0)}|$ and the correction to the quartic evaluated at the unperturbed crossing $\Delta\lambda(\Lambda_\lambda^{(0)})$. The latter is given by eq.~(\ref{eq:del0}) and it remains to identify $|\beta_\lambda^{(0)}|$.

To evaluate the fractional shift in eq.~(\ref{eq:shift}), we must determine the slope of the SM quartic at the unperturbed crossing scale. This is obtained directly from the renormalisation group equation for $\lambda$ (see, for example, eq.~(99) of \cite{Buttazzo:2013uya}). Evaluating the RGE at $\lambda(\Lambda_\lambda^{(0)})=0$ gives, to leading order
\beq
\frac{\dd\lambda}{\dd\ln\bar{\mu}^{ 2}} =\frac{1}{(4\pi)^2}\left[-3y_t^4+\frac{9}{16}g_2^4+\frac{27}{400}g_1^4+\frac{9}{40}g_2^2g_1^2\right],
\label{eq:1loop}
\eeq
with GUT-normalised hypercharge coupling $g_1=\sqrt{\frac{5}{3}} g_Y$. 

The NNLO running couplings at the quartic vanishing scale, $\Lambda_\lambda^{(0)}\sim10^{11}\,\mathrm{GeV}$, are approximately \cite{Buttazzo:2013uya}
\begin{tightequation}
y_t\simeq0.51,\qquad
g_2\simeq0.56,\qquad
g_1\simeq0.53.
\end{tightequation}
\noindent Using the NNLO running couplings in the dominant one-loop contribution, given in eq.~(\ref{eq:1loop}), yields
\beq
\frac{\dd\lambda}{\dd\ln\bar{\mu}^{\,2}}
\simeq
-8\times10^{-4}.
\eeq
Since \cite{Buttazzo:2013uya} defines the RG equation with respect to $\ln\bar{\mu}^{\,2}$, while we define $\beta_\lambda \equiv \frac{\dd\lambda}{\dd\ln Q}$, the two are related by
\beq
\frac{\dd\lambda}{\dd\ln Q} =2\frac{\dd\lambda}{\dd\ln\bar{\mu}^{\,2}}.
\eeq
Thus, after a change of variables, we obtain
\beq
\beta_\lambda^{(0)}
\equiv \left.\frac{\dd\lambda}{\dd\ln Q} \right|_{\Lambda_\lambda^{(0)}}
\simeq -1.6\times10^{-3}.
\label{eq:beta}
\eeq
Combining eqs.~(\ref{eq:del}), (\ref{eq:shift}) and (\ref{eq:beta}) and normalising to the largest consistent hierarchy between the Higgsino LSP and the SUSY scale at $\Lambda_\lambda^{(0)}$ we take $\Lambda_\lambda^{(0)}/m_\chi=10.6$, this implies a fractional shift  
\beq
\frac{\Delta\Lambda_\lambda}{\Lambda_\lambda^{(0)}}
\simeq
8.7\times10^{-3}
\left(\frac{\ln(\Lambda_\lambda^{(0)}/m_\chi)}{\ln 10.6}\right)^2.
\eeq
Thus, the quartic vanishing scale is shifted upward by approximately $0.9\%$.
Since the correction is below one percent, we can identify  $\Lambda_\lambda^{(0)} \simeq \Lambda_\lambda = \widetilde m$.

\section{Higgsino splitting with Dirac gauginos}
\label{ApB}

In this Appendix we derive the neutral-Higgsino mass splitting in the presence of Dirac gaugino masses and $R$-symmetry-breaking Majorana terms. We assume that the gaugino mass eigenstates are heavier than the Higgsinos and sufficiently separated from them that electroweak mixing can be treated perturbatively. To expand the inverse gaugino mass matrix, we require
\beq
\max_{x=\pm\mu}
\left|
\frac{(M_i-x)(M_{A_i}-x)}{M_{D_i}^2}
\right|\ll1,
\label{order}
\eeq
for $i=1,2$, where $M_{D_i}$ is the Dirac gaugino mass, $M_i$ is the gaugino Majorana mass and $M_{A_i}$ the Majorana mass of the adjoint fermion. This condition does not require a particular ordering between $|\mu|$ and $|M_{A_i}|$. As a simplifying assumption we take $\mu$, $M_{D_i}$, $M_i$, and $M_{A_i}$ to be real.

For a single gauge group, the neutralino mass matrix may be written in the basis
$(\lambda_i,\psi_{A_i},\widetilde H_d^0,\widetilde H_u^0)$ as
\beq
\mathcal M_i=
\begin{pmatrix}
M_i & M_{D_i} & -a_i c_\beta & a_i s_\beta\\
M_{D_i} & M_{A_i} & 0 & 0\\
-a_i c_\beta & 0 & 0 & -\mu\\
a_i s_\beta & 0 & -\mu & 0
\end{pmatrix},
\label{MM}
\eeq
where $a_1=m_Zs_W$, and $a_2=m_Zc_W$.

In the limit $a_i=0$ the matrix decomposes into the upper left block $M_{G_i}$ which contains gaugino masses, and the lower right block $M_{H}$ which corresponds to the Higgsinos. In this limit the  Higgsino eigenstates are
\beq
\widetilde H_\pm
= \frac{1}{\sqrt2} \left(\widetilde H_d^0\mp\widetilde H_u^0\right),
\eeq
the mass eigenvalues prior to mixing are
\beq
x_\pm^{(0)}=\pm\mu.
\eeq

We adopt the standard two-component notation
\beq
{\cal L}_{\rm mass}=-\frac12\Psi_i^T{\cal M}_i\Psi_i+{\rm h.c.},
\eeq
where $\Psi_i= (\lambda_i,\psi_{A_i},\widetilde H_d^0,\widetilde H_u^0)^{T}$ is the column vector of neutral gauge-eigenstate Weyl fermions. Since ${\cal M}_i$ is symmetric, an off-diagonal entry ${\cal M}_{ab}$ contributes $-{\cal M}_{ab}\psi_a\psi_b+{\rm h.c.}$ to the mass Lagrangian. The entries 
\beq
{\cal M}_{\lambda_i\widetilde H_d^0}=-a_ic_\beta, \qquad {\cal M}_{\lambda_i\widetilde H_u^0}=a_is_\beta,
\eeq
 give the Higgsino-gaugino mixing terms
\beq
{\cal L}_{\rm mix}
=\lambda_i\left(a_ic_\beta\,\widetilde H_d^0-a_is_\beta\,\widetilde H_u^0\right)+{\rm h.c.}
\label{eq:Lmix}
\eeq
We assume that the adjoint fermion $\psi_{A_i}$ has no direct mixing with the Higgsinos. Expressing the gauge eigenstates in terms of the Higgsino mass eigenstates
\beq
\widetilde H_d^0
&=\frac{1}{\sqrt2}\left(\widetilde H_++\widetilde H_-\right),\\
\widetilde H_u^0
&=
\frac{1}{\sqrt2}\left(\widetilde H_--\widetilde H_+\right),
\eeq
gives
\beq
{\cal L}_{\rm mix}=-\lambda_i\left(m_{i+}\widetilde H_++m_{i-}\widetilde H_-\right)+{\rm h.c.},
\eeq
where
\beq
m_{i+} &=-\frac{a_i}{\sqrt2} (c_\beta+s_\beta),\\
m_{i-} &= \frac{a_i}{\sqrt2} (s_\beta-c_\beta).
\eeq
These quantities are the Higgsino-gaugino mixing matrix elements in the Higgsino mass eigenbasis. Furthermore, their squared magnitudes determine the leading perturbative correction to the Higgsino masses after integrating out the heavy gaugino sector
\beq
|m_{i\pm}|^2=\frac{a_i^2}{2}(1\pm\sin2\beta).
\label{eq:m2}
\eeq
Integrating out the heavy gaugino states gives the leading correction to the mass eigenvalues
\beq
\delta x_{i\pm}=-|m_{i\pm}|^2\left(M_{G_i}-x_\pm^{(0)}\right)^{-1}_{11},
\eeq
where $M_{G_i}$  is the heavy gaugino block of eq.~(\ref{MM}).Since the Higgsino mixing vector in the heavy-sector basis
$(\lambda_i,\psi_{A_i})$ is $(m_{i\pm},0)^T$, contracting it with the
inverse heavy block selects only its $(1,1)$ component. This inverse element can be expressed as
\beq
\left(M_{G_i}-x\right)^{-1}_{11}=\frac{M_{A_i}-x}{(M_i-x)(M_{A_i}-x)-M_{D_i}^2}.
\eeq
Expanding this inverse element under the condition of eq.~(\ref{order}), the correction
reduces to
\beq
\delta x_{i\pm}\simeq \frac{|m_{i\pm}|^2}{M_{D_i}^2}\left(M_{A_i}\mp\mu\right).
\label{eq:hss}
\eeq
The physical masses are obtained from the absolute values of the eigenvalues. For $|\delta x_\pm|\ll|\mu|$, this gives
\beq
m_\pm
\simeq |\mu| \pm {\rm sgn}(\mu)\,\delta x_\pm,
\eeq
where the total shift of each Higgsino eigenvalue is
\beq
\delta x_\pm = \sum_{i=1}^2\delta x_{i\pm}.
\eeq
The resulting physical neutral-Higgsino splitting is
\beq
\delta m_0^{\rm Dirac}
\equiv |m_+-m_-|\simeq\left|\delta x_++\delta x_-\right|.
\eeq
Using  eqs.~(\ref{eq:m2}) and (\ref{eq:hss}) this gives the expression in the main text
\beq
\delta m_0^{\rm Dirac}
&\simeq m_Z^2 \left|s_W^2 \frac{M_{A_1}-\mu\sin2\beta}{M_{D_1}^2}
+c_W^2 \frac{M_{A_2}-\mu\sin2\beta}{M_{D_2}^2}
\right|.
\notag 
\eeq



\end{document}